\documentclass{appolb}
\usepackage{epsfig,dcolumn}
\usepackage{graphicx}
\usepackage{color}

\usepackage{amsmath}
\usepackage{amsfonts}
\usepackage{amssymb}
\usepackage{graphicx}
\usepackage{type1cm}
\usepackage{eso-pic}
\newcommand{\be}{\begin{equation}}
\newcommand{\ee}{\end{equation}}

\newcommand{\im}{\mathrm{Im}\,}
\newcommand{\re}{\mathrm{Re}\,}

\usepackage{bm} % scaled bold math symbols

\begin{document}

\title{Forward dispersion relations for $\pi$-K scattering and the $K^*_0(800)$ resonance%
\thanks{Presented at Excited QCD 2016, Costa do Caparica, Portugal, March 6-12}%
}
\author{A.Rodas, J.R.Pel\'aez
\address{Departamento de F\'isica Te\'orica II, Universidad Complutense de Madrid, 28040 Madrid, Spain}
\\
}
\maketitle
\begin{abstract}
We review our recent analysis of $\pi K$ scattering data in terms of forward dispersion relations, and also present the parameters of the strange resonances. This work consists of fits to the data that are constrained to satisfy analyticity requirements. The method yields a set of simple and consistent parameterizations that are compatible with forward dispersion relations up to 1.6 GeV while still describing the data. We also obtain the pole parameters of the $K^*_0(800)$ and the $K^*(892)$ resonances.
\end{abstract}
  
\section{Introduction}

A precise knowledge of $\pi K$ scattering is of interest since it is one of the main reactions involved in the final state of hadronic processes with net strangeness. It is also interesting by itself because it provides a test of Chiral Perturbation Theory \cite{Gasser:1984gg} and other unitarized approaches \cite{Dobado:1992ha}. Moreover, there is a renewed interest in $\pi K$ scattering 
from Lattice QCD, where the 
main features, like threshold parameters \cite{Beane:2006gj},
scattering phases and resonances \cite{Lang:2012sv}, have already been calculated. Finally, our study of the scalar channel leads to a clear pole for the  still controversial $K^*_0(800), \kappa$, which according to the PDG still needs confirmation.

From the experimental point of view this processes cannot be directly measured and hence the data is plagued by systematic uncertainties. The data available in the bibliography \cite{Estabrooks:1977xe} are clearly incompatible, furthermore, there are no data close to the threshold.

Our goal is to perform an analysis of this scattering process using only analytic constraints and data. The dispersive integral formalism is model independent and relates the value of the amplitude with an integral over the whole real axis, increasing the precision and giving information even in regions where data are lacking or have large uncertainties. In addition, it relates different channels among themselves. Moreover it is also useful
to constrain threshold and resonance parameters.

Our work \cite{Pelaez:2016tgi} is based on Forward Dispersion Relations (FDR). As they are calculated at $t=0$ we can use this set of equations up to arbitrary energies in the real axis, providing a set of simple but powerful constraints for the fits. We consider two independent amplitudes, one symmetric and one antisymmetric under the $s \leftrightarrow u$ exchange that cover the isospin basis $T^+(s)=(T^{1/2}(s)+2T^{3/2}(s))/3=T^{I_t=0}(s)/\sqrt{6}$ and $T^-(s)=(T^{1/2}(s)-T^{3/2}(s))/3=T^{I_t=1}(s)/2$.. The symmetric has one subtraction and can be written as

\begin{align}
&\re T^+(s)=T^+(s_{th})+\frac{(s-s_{th})}{\pi}\times \nonumber \\
&\times P\!\!\!\int^{\infty}_{s_{th}}\!\!\!\! ds'\!\left[\!\frac{\im T^+(s')}{(s'-s)(s'-s_{th})} 
-\frac{\im T^+(s')}{(s'+s-2\Sigma_{\pi K})(s'+s_{th}-2\Sigma_{\pi K})}\!\right],
\label{eq:FDRTsym}
\end{align}

where $s_{th}=(m_\pi+m_K)^2$.
In contrast the antisymmetric one does not require subtractions:

\begin{equation}
\re T^-(s)=
\frac{(2s-2\Sigma_{\pi K})}{\pi}P\!\!\int^{\infty}_{s_{th}}\!\!\!\! ds'
\frac{\im T^-(s')}{(s'-s)(s'+s-2\Sigma_{\pi K})}.
\label{eq:FDRTan}
\end{equation}

We also include in our analysis 3 sum rules for threshold parameters (scattering lengths and slopes) in order to obtain the best possible result in this region, where there are no data. These integral equations allow us to obtain a set of precise scattering lengths. There can be observed \cite{Pelaez:2016tgi} how this analysis leads to a set of threshold parameters compatible with the experimental measurements of the DIRAC collaboration \cite{Adeva:2014xtx}.

There exists previous dispersive analysis \cite{DescotesGenon:2006uk} that obtains the phase shifts of the scalar and vectorial channels by using Roy-Steiner equations (RS). But these equations can only be applied in the low energy region ($\sqrt{s}\leq 0.935$ GeV).

\section{Method and results}

The approach used in this work follows the same steps as previous works done by our group for $\pi-\pi$ scattering \cite{GarciaMartin:2011cn}. 
(1) We first obtain simple fits for each partial wave independently, called Unconstrained Fits to Data (UFD), without including any model description. (2) We check the fulfillment of the dispersion relations to observe if there are some inconsistent data points that do not satisfy it. (3) We finally impose this integral equations to obtain the final Constrained Fits to Data (CFD), where all the partial waves are related through the FDR and the sum rules.

\begin{figure}
\centering
\centerline{\includegraphics[width=0.42\linewidth]{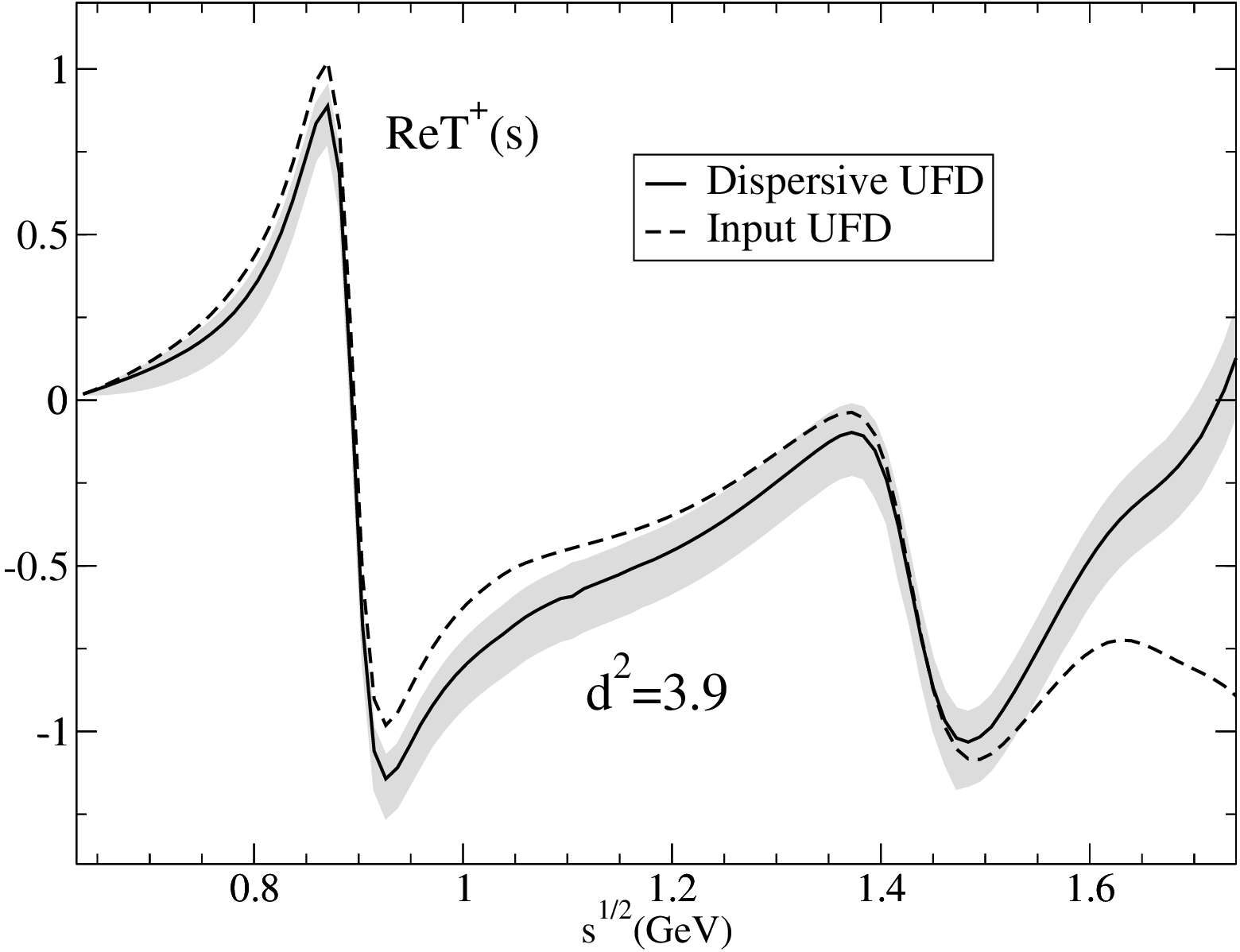}   \includegraphics[width=0.42\linewidth]{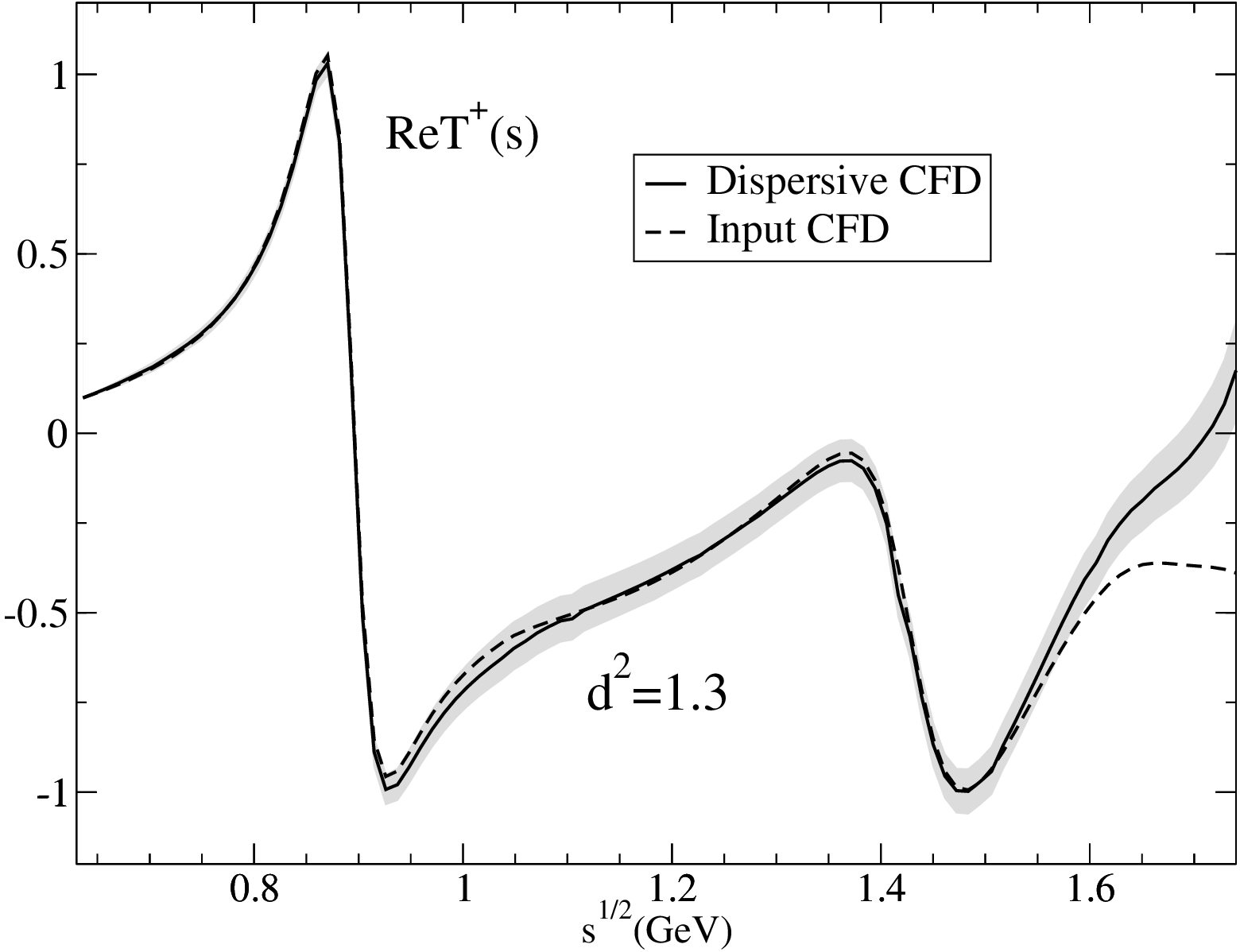}}
\vspace{0.15 cm}
\centerline{\includegraphics[width=0.42\linewidth]{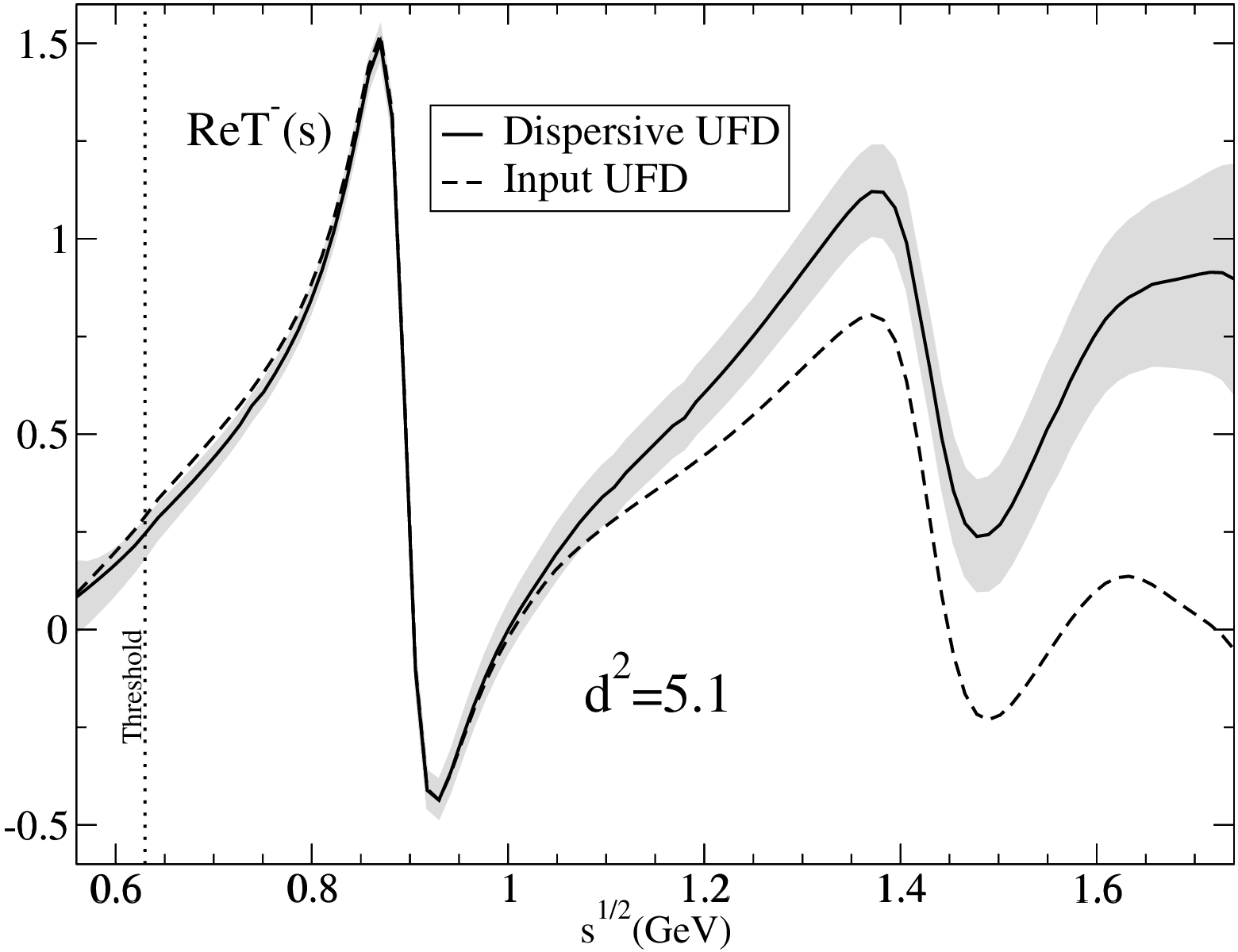} \includegraphics[width=0.42\linewidth]{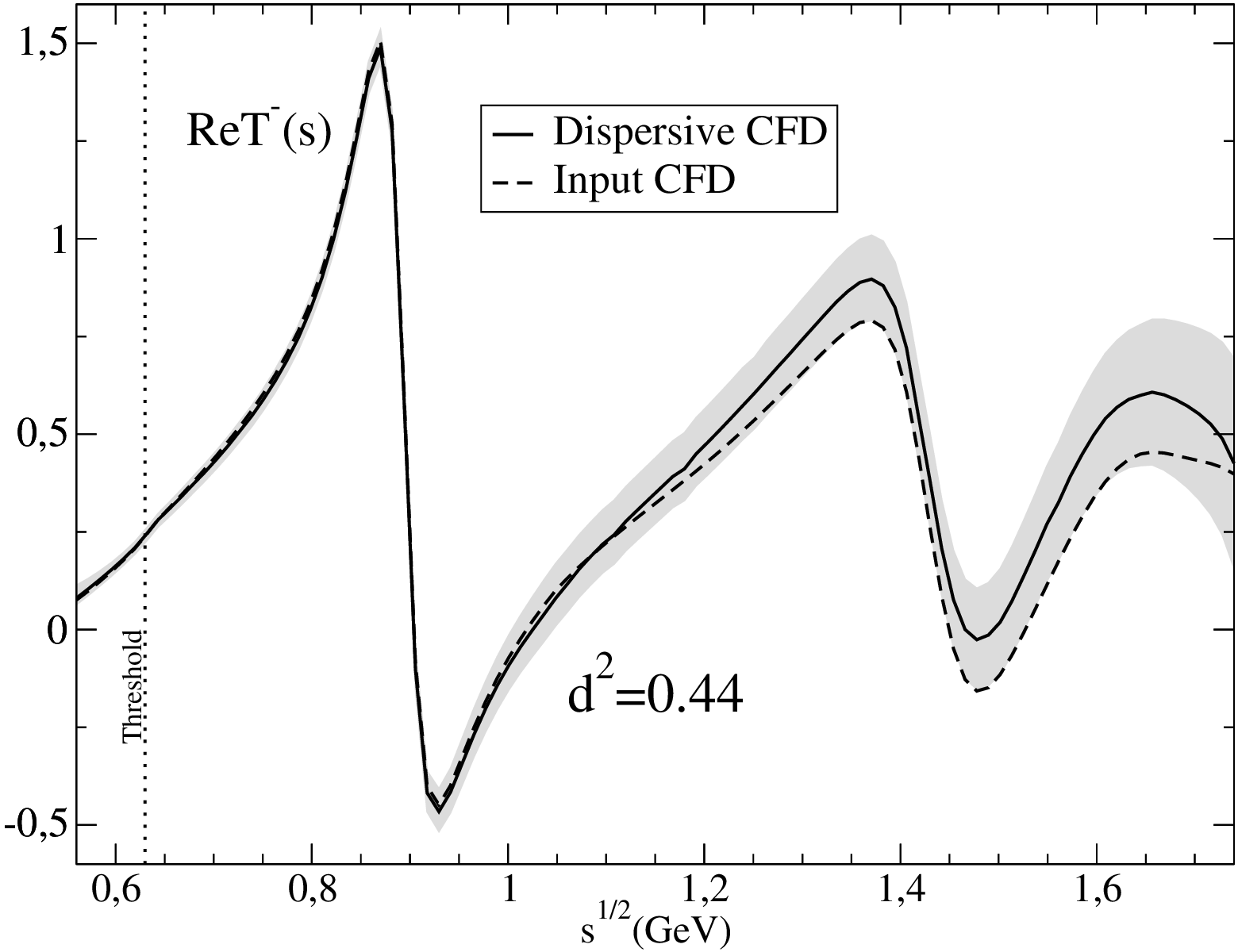}}
\vspace{-0.1 cm}
\caption{\rm \label{fig:FDR} 
Comparison between the input (fits)
and the output(FDRs) for the total amplitudes $T^+$ (top) and $T^-$(bottom).
The gray bands describe the uncertainty of the difference between the input and the output.
}
\end{figure}

In order to impose the FDRs we define $d_i$ the difference between the input and the output of each dispersion relation at the energy point $s_i$, whose uncertainties are $\Delta d_i$. We thus define the average discrepancies

\begin{equation} 
d_{T\pm}^2=\frac{1}{N}\sum_{i=1}^{N} \left(\frac{d_i}{\Delta d_i}\right)_{T^\pm}^2.
\label{eq:distances}
\end{equation}

We introduce a penalty function that measures the difference between the UFD parameters and the CFD ones to describe also the data, obtaining the final $\chi^2$ function

\begin{equation}
\chi^2=W^2(d_{T^+}^2+d_{T^-}^2)+d_{SR}
+\sum_k \left (\frac{p_k^{UFD}-p_k}{\delta p_k^{UFD}}\right)^2.
\end{equation}

The weight $W^2=12$ stands for the FDR that is just the number of degrees of freedom needed to describe the amplitude in the region of interest.

\begin{figure}
\centering
\centerline{\includegraphics[width=0.44\linewidth]{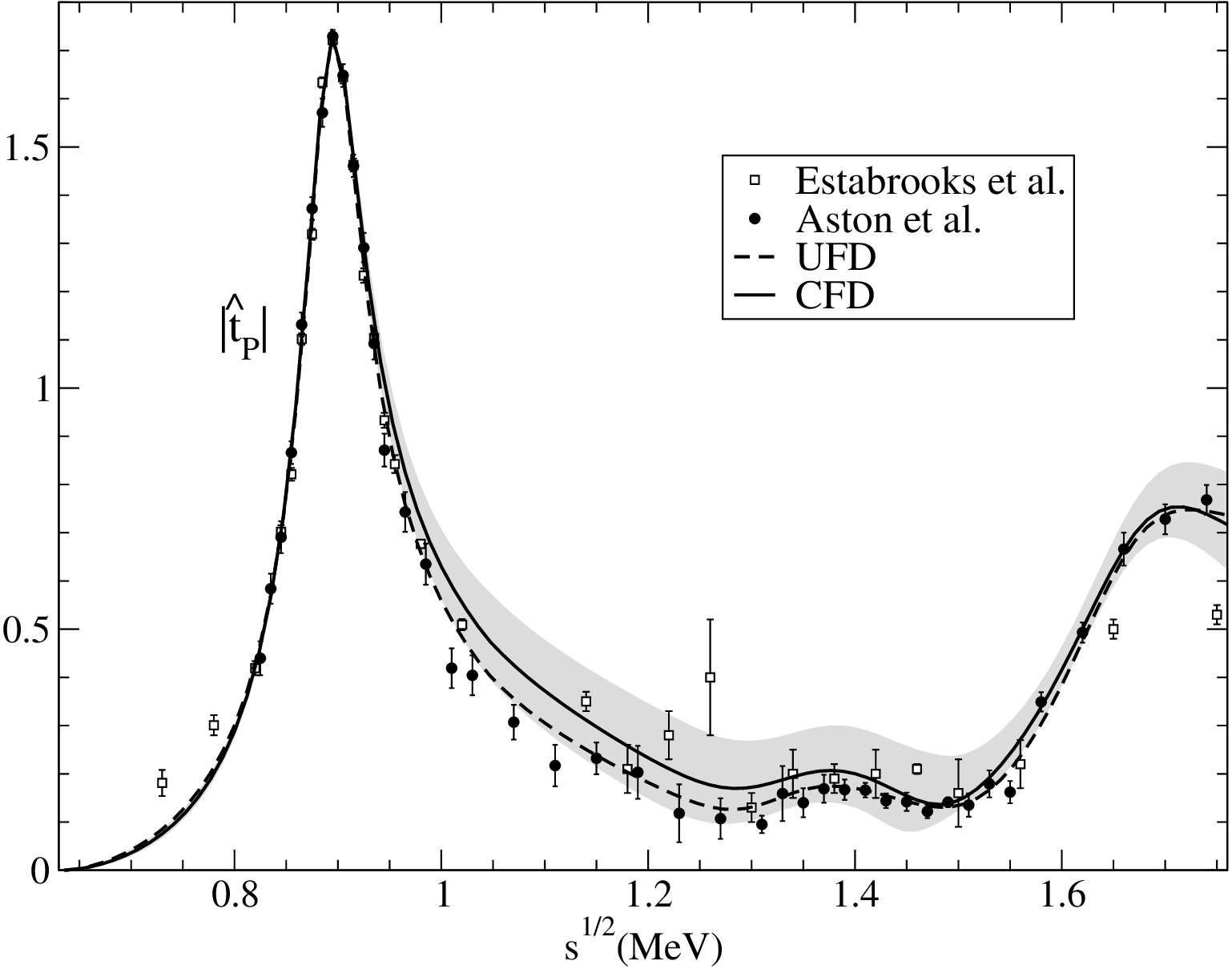} \hspace{.05cm}  \includegraphics[width=0.44\linewidth]{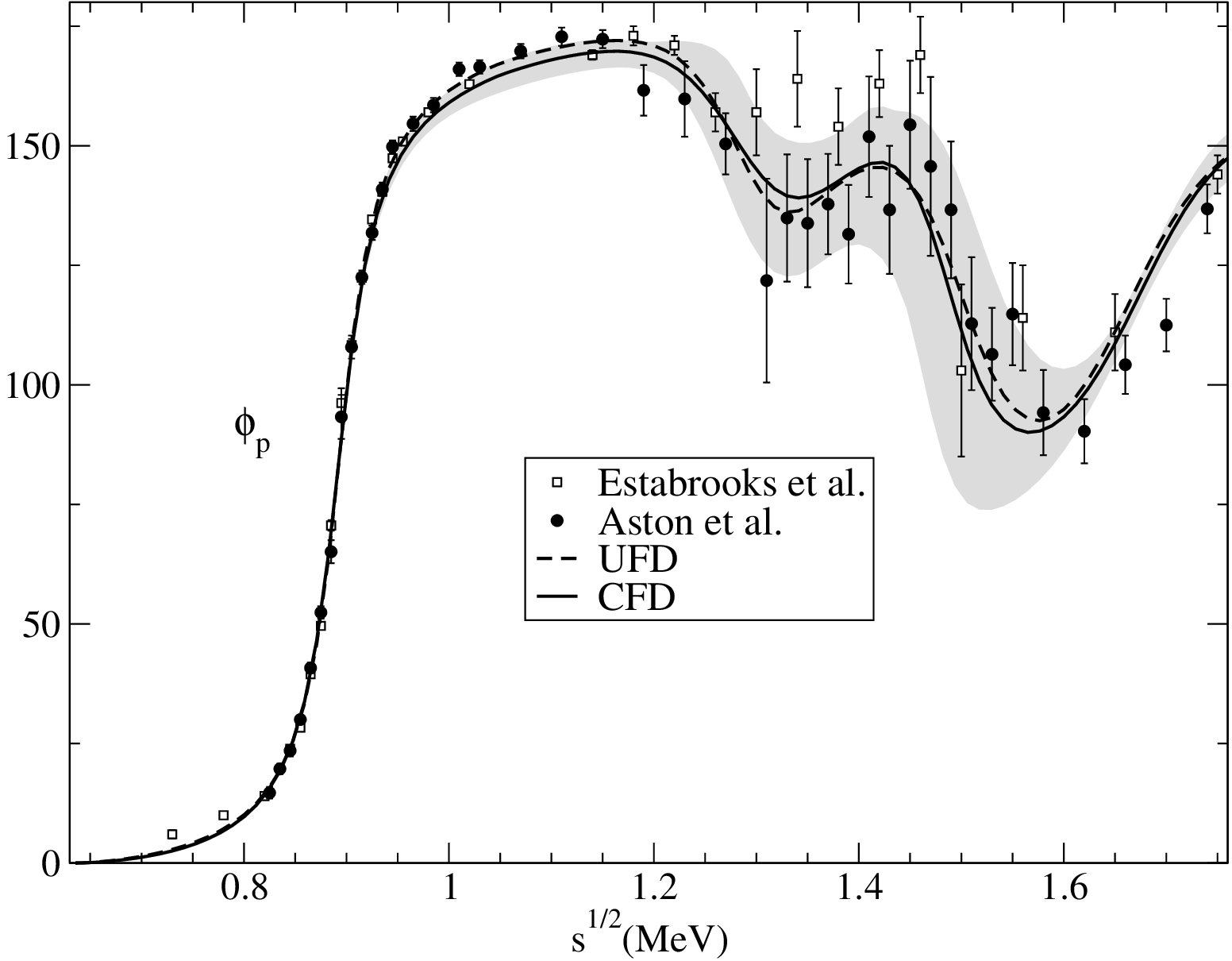}}
\centerline{\includegraphics[width=0.42\linewidth]{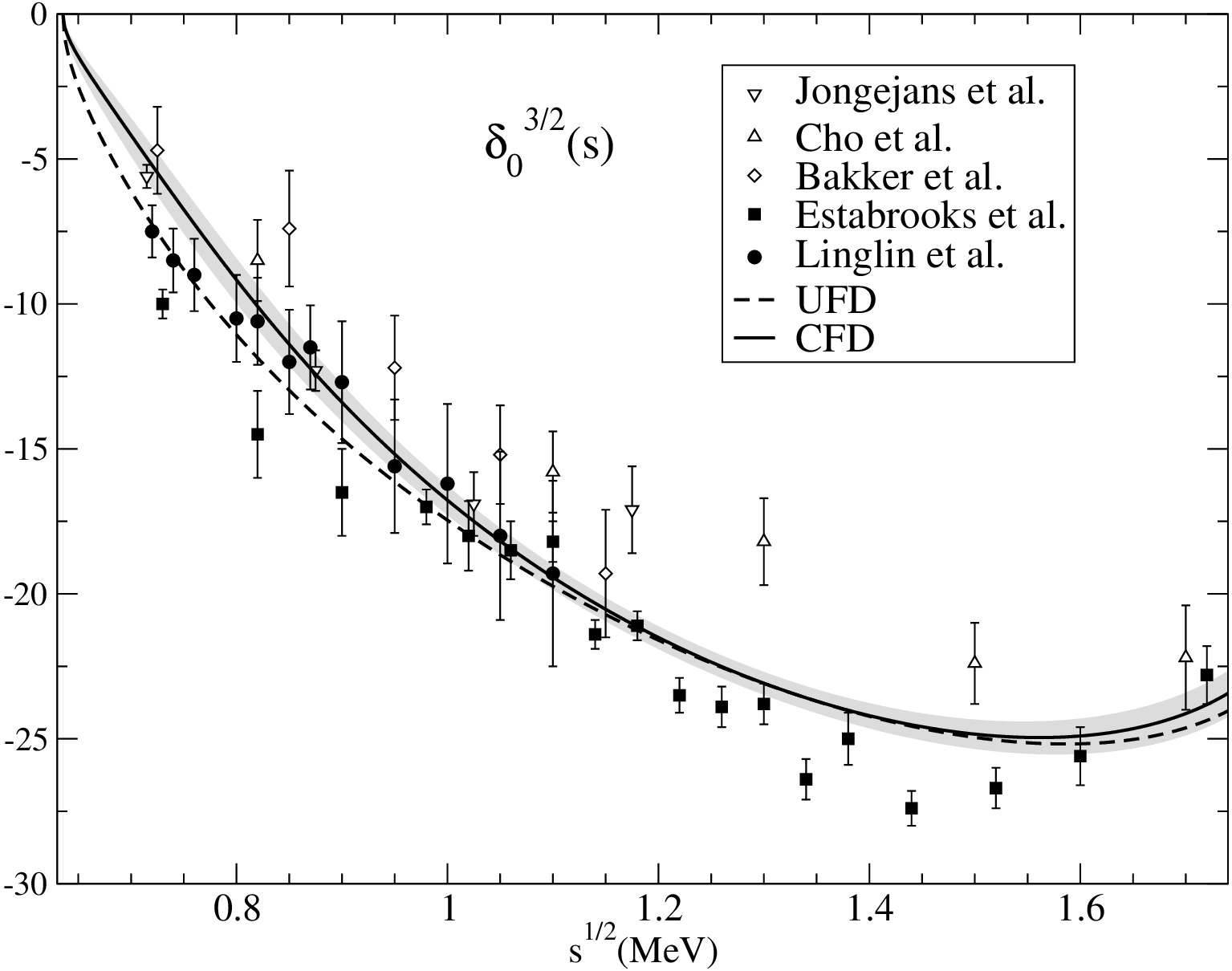} \includegraphics[width=0.5\linewidth]{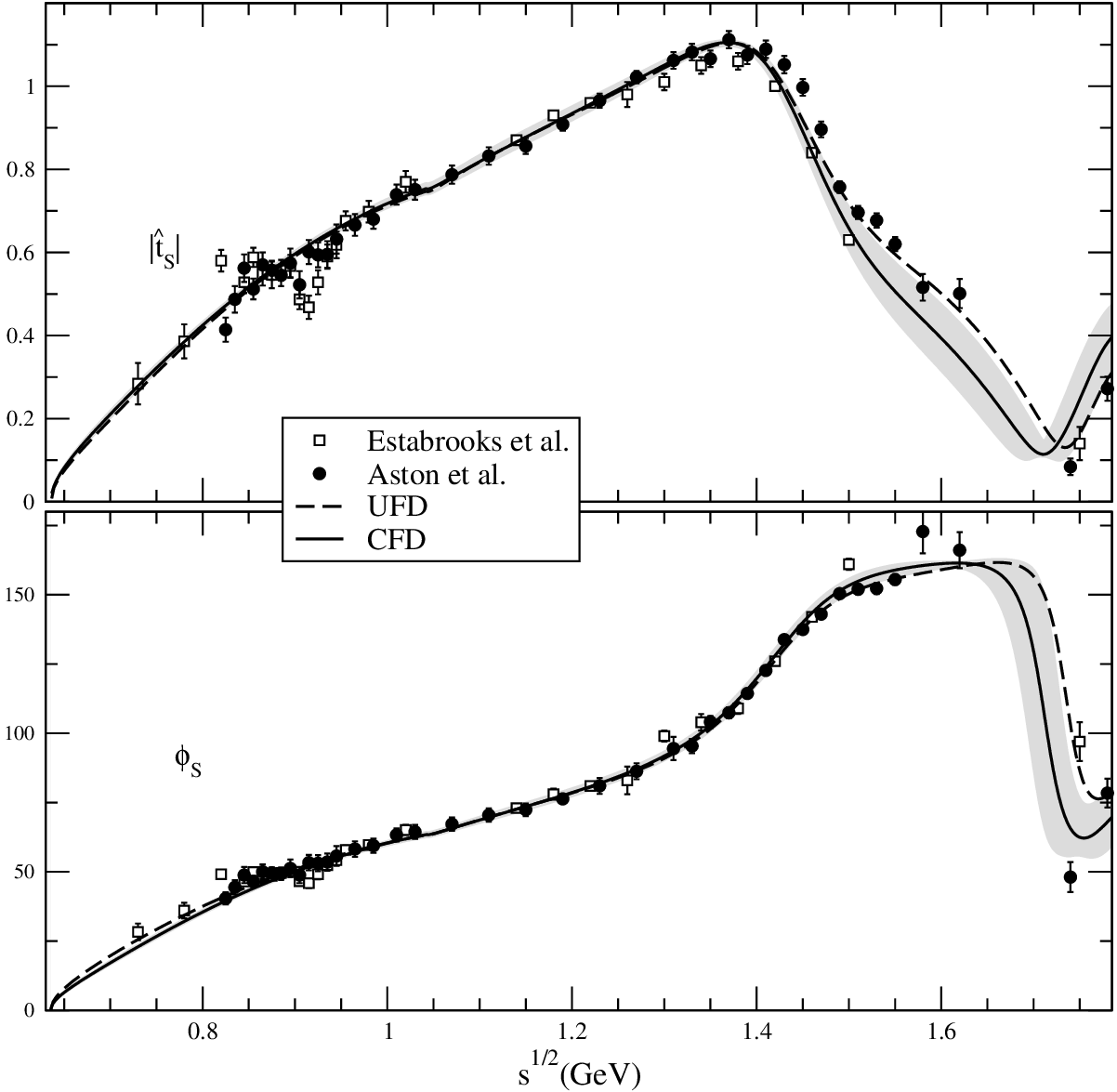}}
\caption{\rm \label{fig:CFD} 
Comparison between UFD and CFD fits for different partial waves, where $|\hat{t}|$ stands for the modulus, $\delta$ for the phase shift and $\phi$ for the total phase of each partial wave. The gray bands cover the errors of the parameters for each fit.
}
\end{figure}

Fig.\ref{fig:FDR} shows the total amplitudes and the huge improvement between the UFD and the CFD parametrizations, in Fig.\ref{fig:CFD} we show the difference between the fits to the data and the final results. 

\begin{table} 
\caption{Fulfillment of Forward Dispersion Relations.} 
\centering 
\begin{tabular}{c | c c | c c} 
\hline\hline  
\rule[-0.15cm]{0cm}{.55cm} & \multicolumn{2}{c|}{UFD} & \multicolumn{2}{c}{CFD}\\ 
\rule[-0.15cm]{0cm}{.55cm}& $d_{T^+}^2$ & $d_{T^-}^2$& $d_{T^+}^2$ & $d_{T^-}^2$\\
\hline 
\rule[-0.15cm]{0cm}{.55cm} $\sqrt{s_{min}}\leq\sqrt{s}\leq m_K+m_\eta$&3.35  & 0.97& 0.39& 0.13\\
\rule[-0.15cm]{0cm}{.55cm} $m_K+m_\eta \leq\sqrt{s}\leq 1.6\,$GeV& 1.3& 6.8& 0.17& 0.70\\
\rule[-0.15cm]{0cm}{.55cm} $1.6\,$GeV$\leq\sqrt{s}\leq 1.74\,$GeV& 14.6& 12.8& 8.0& 0.5\\
\hline 
\rule[-0.15cm]{0cm}{.55cm} $\sqrt{s_{min}}\leq\sqrt{s}\leq 1.74\,$GeV& 3.9& 5.1& 1.3& 0.44\\
\hline 
\end{tabular} 
\label{tab:d2FDR} 
\end{table} 

We show in Table~\ref{tab:d2FDR} the average discrepancies for different energy regions. As it can be observed the CFD result is very consistent below 1.6 GeV.

Now we use the CFD parameters to give the result for the most important threshold parameters, obtaining $m_\pi a_0^{1/2}=0.22\pm 0.01$ and $m_\pi a_0^{3/2}=-0.054^{+0.010}_{-0.014}$. We show here our results compared with the recent result measured by the Dirac collaboration

\begin{equation}
\frac{1}{3}\left(a_0^{1/2}-a_0^{3/2}\right)=0.11^{+0.09}_{-0.04} \, m_\pi^{-1},\quad ({\rm DIRAC})
\label{eq:dirac}
\end{equation}

\begin{equation}
\frac{1}{3}\left(a_0^{1/2}-a_0^{3/2}\right)=0.091^{+0.006}_{-0.005} \, m_{\pi}^{-1}.\quad ({\rm CFD})
\label{ec:DIRACUFD}
\end{equation}

Using our conformal parametrization we can also continue the partial waves to the complex plane. Calculating the position of the resonances in the second Riemann sheet. In Tables~\ref{tab:polesK0} and \ref{tab:polesKstar} we show the mass, width and coupling of each elastic resonance, defined as $s_{R}=(M_R-i \Gamma_R/2)^2$ and $\vert g \vert^2=\vert 16 \pi (2l+1)\,{\rm Res}(t_l(s_R))/(2q(s_R))^{2l} \vert$.

\begin{table}[h] 
\caption{$K^*_0(800)$ pole parameters
from the analytic continuation of the elastic parameterization.} 
\centering 
\begin{tabular}{c c c c} 
\hline\hline  
Poles & Mass (MeV) & Width (MeV) & $\vert g \vert$ (GeV)\\ 
\hline 
$UFD$    &  673$\pm$15  &  674$\pm$15 & 5.01$\pm$0.07 \\
$CFD$    &  680$\pm$15  &  668$\pm$15 & 4.99$\pm$0.08 \\
\hline 
\end{tabular} 
\label{tab:polesK0} 
\end{table}

\begin{table}[h] 
\caption{$K^*(892)$ pole parameters
from the analytic continuation of the elastic parameterization only.} 
\centering 
\begin{tabular}{c c c c} 
\hline\hline  
Poles & Mass (MeV) & Width (MeV) & $\vert g \vert$ \\ 
\hline 
$UFD$    &  893$\pm$1  &  56$\pm$2 & 5.95$\pm$0.07 \\
$CFD$    &  892$\pm$1 &  58$\pm$2  & 6.02$\pm$0.06 \\
\hline 
\end{tabular} 
\label{tab:polesKstar} 
\end{table}

\section{Outloook}

Fig.\ref{fig:FDR} shows that the CFD set satisfies really well the dispersion relations up to 1.6 GeV. Above that energy the differences between the input and the output require larger deviations from  data as it is shown in Fig.\ref{fig:CFD}.

Finally the use of conformal mappings in the elastic region allows us to continue the partial waves to the second Riemann sheet. The values obtained for the parameters of the resonances are in agreement with other works listed in the PDG, although we obtain smaller uncertainties due to the small error of the CFD parameters.

\section{Acknowledgements}
J.R.P. and A.R. are supported by Spanish Projects No. FPA2011-27853-C02-02 and No. FPA2014-53375C2-2 and Red de Excelencia de Física Hadrónica FIS2014-57026-REDT.


\begin{thebibliography}{99}



%\cite{Gasser:1984gg}
\bibitem{Gasser:1984gg} 
  J.~Gasser and H.~Leutwyler,
  %``Chiral Perturbation Theory: Expansions in the Mass of the Strange Quark,''
  Nucl.\ Phys.\ B {\bf 250}, 465 (1985).
  %%CITATION = NUPHA,B250,465;%%
  %3170 citations counted in INSPIRE as of 23 Mar 2015
  V.~Bernard, N.~Kaiser and U.~G.~Meissner,
  %``pi K scattering in chiral perturbation theory to one loop,''
  Nucl.\ Phys.\ B {\bf 357}, 129 (1991).
%  doi:10.1016/0550-3213(91)90461-6
  %%CITATION = doi:10.1016/0550-3213(91)90461-6;%%
  %138 citations counted in INSPIRE as of 25 janv. 2016



%\cite{Dobado:1992ha}
\bibitem{Dobado:1992ha} 
  A.~Dobado and J.~R.~Pelaez,
  %``A Global fit of pi pi and pi K elastic scattering in ChPT with dispersion relations,''
  Phys.\ Rev.\ D {\bf 47}, 4883 (1993).
%  doi:10.1103/PhysRevD.47.4883
%  [hep-ph/9301276].
  %%CITATION = doi:10.1103/PhysRevD.47.4883;%%
  %191 citations counted in INSPIRE as of 25 janv. 2016
%\cite{GomezNicola:2001as}
%\bibitem{GomezNicola:2001as} 
  A.~Gomez Nicola and J.~R.~Pelaez,
  %``Meson meson scattering within one loop chiral perturbation theory and its unitarization,''
  Phys.\ Rev.\ D {\bf 65}, 054009 (2002).
%  doi:10.1103/PhysRevD.65.054009
%  [hep-ph/0109056].
  %%CITATION = doi:10.1103/PhysRevD.65.054009;%%
  %184 citations counted in INSPIRE as of 25 janv. 2016
  J.~R.~Pelaez,
  %``Light scalars as tetraquarks or two-meson states from large N(c) and unitarized chiral perturbation theory,''
  Mod.\ Phys.\ Lett.\ A {\bf 19}, 2879 (2004).
%  doi:10.1142/S0217732304016160
%  [hep-ph/0411107].
  %%CITATION = doi:10.1142/S0217732304016160;%%
  %113 citations counted in INSPIRE as of 25 janv. 2016
  J.~A.~Oller and E.~Oset,
  %``Chiral symmetry amplitudes in the S-wave isoscalar and isovector  channels
  %and the sigma, f0(980), a0(980) scalar mesons,''
  Nucl.\ Phys.\ A {\bf 620}, 438 (1997)
  [Erratum-ibid.\ A {\bf 652}, 407 (1999)].
%  [arXiv:hep-ph/9702314].
  %%CITATION = HEP-PH 9702314;%%
%
%\cite{Oller:1998zr}
%\bibitem{Oller:1998zr}
%J.~A.~Oller and E.~Oset,
  %``N/D description of two meson amplitudes and chiral symmetry,''
  Phys.\ Rev.\ D {\bf 60} (1999) 074023.
%  [hep-ph/9809337].
  %%CITATION = HEP-PH/9809337;%%
%\cite{Oller:1997ng}
%\bibitem{Oller:1997ng}
J.~A.~Oller, E.~Oset and J.~R.~Pelaez,
%``Non-perturbative approach to effective chiral Lagrangians and meson  interactions,''
Phys.\ Rev.\ Lett.\  {\bf 80} (1998) 3452,
%[arXiv:hep-ph/9803242].
%%CITATION = HEP-PH 9803242;%%
%\cite{Oller:1998hw}
%\bibitem{Oller:1998hw} 
%  J.~A.~Oller, E.~Oset and J.~R.~Pelaez,
  %``Meson meson interaction in a nonperturbative chiral approach,''
  Phys.\ Rev.\ D {\bf 59}, 074001 (1999)
  [Erratum-ibid.\ D {\bf 60}, 099906 (1999)]
  [Erratum-ibid.\ D {\bf 75}, 099903 (2007)].
%  [hep-ph/9804209].
  %%CITATION = HEP-PH/9804209;%%
  %472 citations counted in INSPIRE as of 31 Oct 2014




%\cite{Beane:2006gj}
\bibitem{Beane:2006gj} 
  S.~R.~Beane, {\it et al. }%P.~F.~Bedaque, T.~C.~Luu, K.~Orginos, E.~Pallante, A.~Parreno and M.~J.~Savage,
  %``pi K scattering in full QCD with domain-wall valence quarks,''
  Phys.\ Rev.\ D {\bf 74}, 114503 (2006).
%  doi:10.1103/PhysRevD.74.114503
%  [hep-lat/0607036].
  %%CITATION = doi:10.1103/PhysRevD.74.114503;%%
  %73 citations counted in INSPIRE as of 26 janv. 2016
%\cite{Nagata:2008wk}
%\bibitem{Nagata:2008wk} 
  J.~Nagata, S.~Muroya and A.~Nakamura,
  %``Lattice study of K pi scattering in I = 3/2 and 1/2,''
  Phys.\ Rev.\ C {\bf 80}, 045203 (2009)
  [Phys.\ Rev.\ C {\bf 84}, 019904 (2011)].
%  doi:10.1103/PhysRevC.84.019904, 10.1103/PhysRevC.80.045203
%  [arXiv:0812.1753 [hep-lat]].
  %%CITATION = doi:10.1103/PhysRevC.84.019904, 10.1103/PhysRevC.80.045203;%%
  %18 citations counted in INSPIRE as of 26 Jan 2016
%\cite{Fu:2011wc}
%\bibitem{Fu:2011wc} 
  Z.~Fu,
  %``Lattice study on $\pi K $ scattering with moving wall source,''
  Phys.\ Rev.\ D {\bf 85}, 074501 (2012).
%  doi:10.1103/PhysRevD.85.074501
%  [arXiv:1110.1422 [hep-lat]].
  %%CITATION = doi:10.1103/PhysRevD.85.074501;%%
  %25 citations counted in INSPIRE as of 26 Jan 2016
%\cite{Sasaki:2013vxa}
%\bibitem{Sasaki:2013vxa} 
  K.~Sasaki {\it et al.} [PACS-CS Collaboration],
  %``Scattering lengths for two pseudoscalar meson systems,''
  Phys.\ Rev.\ D {\bf 89}, no. 5, 054502 (2014).
%  doi:10.1103/PhysRevD.89.054502
%  [arXiv:1311.7226 [hep-lat]].
  %%CITATION = doi:10.1103/PhysRevD.89.054502;%%
  %9 citations counted in INSPIRE as of 26 Jan 2016



%\cite{Lang:2012sv}
\bibitem{Lang:2012sv} 
  C.~B.~Lang, L.~Leskovec, D.~Mohler and S.~Prelovsek,
  %``K pi scattering for isospin 1/2 and 3/2 in lattice QCD,''
  Phys.\ Rev.\ D {\bf 86}, 054508 (2012);
%  doi:10.1103/PhysRevD.86.054508
%  [arXiv:1207.3204 [hep-lat]].
  %%CITATION = doi:10.1103/PhysRevD.86.054508;%%
  %42 citations counted in INSPIRE as of 26 Jan 2016
%\cite{Prelovsek:2013ela}
%\bibitem{Prelovsek:2013ela}
%  S.~Prelovsek, L.~Leskovec, C.~B.~Lang and D.~Mohler,
  %``K $\pi$ scattering and the K* decay width from lattice QCD,''
  Phys.\ Rev.\ D {\bf 88} (2013) 5,  054508.
%  doi:10.1103/PhysRevD.88.054508
%  [arXiv:1307.0736 [hep-lat]].
  %%CITATION = doi:10.1103/PhysRevD.88.054508;%%
  %28 citations counted in INSPIRE as of 26 Jan 2016
%\cite{Fu:2012tj}
%\bibitem{Fu:2012tj} 
  Z.~Fu and K.~Fu,
  %``Lattice QCD study on $K^\ast(892)$ meson decay width,''
  Phys.\ Rev.\ D {\bf 86}, 094507 (2012).
%  doi:10.1103/PhysRevD.86.094507
%  [arXiv:1209.0350 [hep-lat]].
  %%CITATION = doi:10.1103/PhysRevD.86.094507;%%
  %16 citations counted in INSPIRE as of 26 janv. 2016
%\cite{Dudek:2014qha}
%\bibitem{Dudek:2014qha} 
  J.~J.~Dudek {\it et al.} [Hadron Spectrum Collaboration],
  %``Resonances in coupled $\pi K -\eta K$ scattering from quantum chromodynamics,''
  Phys.\ Rev.\ Lett.\  {\bf 113}, 182001 (2014).
%  doi:10.1103/PhysRevLett.113.182001
%  [arXiv:1406.4158 [hep-ph]].
  %%CITATION = doi:10.1103/PhysRevLett.113.182001;%%
  %45 citations counted in INSPIRE as of 25 Jan 2016
%\cite{Wilson:2014cna}
%\bibitem{Wilson:2014cna} 
  D.~J.~Wilson, J.~J.~Dudek, R.~G.~Edwards and C.~E.~Thomas,
  %``Resonances in coupled $\pi K, \eta K$ scattering from lattice QCD,''
  Phys.\ Rev.\ D {\bf 91}, 054008 (2015).
%  doi:10.1103/PhysRevD.91.054008
%  [arXiv:1411.2004 [hep-ph]].
  %%CITATION = doi:10.1103/PhysRevD.91.054008;%%
  %20 citations counted in INSPIRE as of 25 janv. 2016





	


%\cite{Estabrooks:1977xe}
\bibitem{Estabrooks:1977xe}
  P.~Estabrooks {\it et al.}, %R.~K.~Carnegie, A.~D.~Martin, W.~M.~Dunwoodie, T.~A.~Lasinski and D.~W.~G.~S.~Leith,
  %``Study of K pi Scattering Using the Reactions K+- p ---> K+- pi+ n and K+- p ---> K+- pi- Delta++ at 13-GeV/c,''
  Nucl.\ Phys.\ B {\bf 133}, 490 (1978).
  %%CITATION = NUPHA,B133,490;%%
  %254 citations counted in INSPIRE as of 17 sept. 2015
  D.~Aston {\it et al.},
  %``A Study of K- pi+ Scattering in the Reaction K- p ---> K- pi+ n at 11-GeV/c,''
  Nucl.\ Phys.\ B {\bf 296}, 493 (1988).
  %%CITATION = NUPHA,B296,493;%%
  %519 citations counted in INSPIRE as of 17 sept. 2015
  Y.~Cho {\it et al.},
  %``Study of k- pi- scattering using the reaction k- d ---> k- pi- p p(s),''
  Phys.\ Lett.\ B {\bf 32}, 409 (1970).
  %%CITATION = PHLTA,B32,409;%%
  %22 citations counted in INSPIRE as of 17 sept. 2015
  A.~M.~Bakker {\it et al.},
  %``A determination of the i=3/2 k pi elastic-scattering cross section from the reaction k- n ---> p k- pi- at 3 gev/c,''
  Nucl.\ Phys.\ B {\bf 24}, 211 (1970).
  %%CITATION = NUPHA,B24,211;%%
  %27 citations counted in INSPIRE as of 17 sept. 2015
  D.~Linglin {\it et al.},
  %``K- pi- elastic scattering cross-section measured in 14.3 gev/c k- p interactions,''
  Nucl.\ Phys.\ B {\bf 57}, 64 (1973).
  %%CITATION = NUPHA,B57,64;%%
  %29 citations counted in INSPIRE as of 17 sept. 2015
  B.~Jongejans, R.~A.~van Meurs, A.~G.~Tenner, H.~Voorthuis, P.~M.~Heinen, W.~J.~Metzger, H.~G.~J.~M.~Tiecke and R.~T.~Van de Walle,
  %``Study Of The I = 3/2 K- Pi- Elastic Scattering From The Reaction K- P ---> K- Pi- P Pi+ At 4.25-gev/c Incident K- Momentum,''
  Nucl.\ Phys.\ B {\bf 67}, 381 (1973).
  %%CITATION = NUPHA,B67,381;%%
  %22 citations counted in INSPIRE as of 17 sept. 2015



%\cite{Pelaez:2016tgi}
\bibitem{Pelaez:2016tgi} 
  J.~R.~Pelaez and A.~Rodas,
  %``Pion-kaon scattering amplitude constrained with forward dispersion relations up to 1.6 GeV,''
  Phys.\ Rev.\ D {\bf 93}, no. 7, 074025 (2016)
  %doi:10.1103/PhysRevD.93.074025
  %[arXiv:1602.08404 [hep-ph]].
  %%CITATION = doi:10.1103/PhysRevD.93.074025;%%
	

%Dirac group
%\cite{Adeva:2014xtx}
\bibitem{Adeva:2014xtx}
  B.~Adeva {\it et al.} [DIRAC Collaboration],
  %``First ~K atom lifetime and ~K scattering length measurements,''
  Phys.\ Lett.\ B {\bf 735}, 288 (2014).
%  [arXiv:1403.0845 [nucl-ex]].
  %%CITATION = ARXIV:1403.0845;%%
  %6 citations counted in INSPIRE as of 17 sept. 2015
	
	

%\cite{DescotesGenon:2006uk}
\bibitem{DescotesGenon:2006uk} 
  S.~Descotes-Genon and B.~Moussallam,
  %``The K*0 (800) scalar resonance from Roy-Steiner representations of pi K scattering,''
  Eur.\ Phys.\ J.\ C {\bf 48}, 553 (2006).
%  [hep-ph/0607133].
  %%CITATION = HEP-PH/0607133;%%
  %134 citations counted in INSPIRE as of 23 Mar 2015
  P.~B{\"u}ttiker, S.~Descotes-Genon and B.~Moussallam,
  %``A new analysis of pi K scattering from Roy and Steiner type equations,''
  Eur.\ Phys.\ J.\ C {\bf 33}, 409 (2004).
%  [hep-ph/0310283].
  %%CITATION = HEP-PH/0310283;%%
  %160 citations counted in INSPIRE as of 23 Mar 2015 
  B.~Ananthanarayan, P.~B{\"u}ttiker and B.~Moussallam,
  %``pi K sum rules and the SU(3) chiral expansion,''
  Eur.\ Phys.\ J.\ C {\bf 22}, 133 (2001).
%  doi:10.1007/s100520100766
%  [hep-ph/0106230].
  %%CITATION = doi:10.1007/s100520100766;%%
  %38 citations counted in INSPIRE as of 25 Mar 2016
  %\cite{Ananthanarayan:2000cp}
%\bibitem{Ananthanarayan:2000cp} 
  B.~Ananthanarayan and P.~B{\"u}ttiker,
  %``Comparison of pion kaon scattering in SU(3) chiral perturbation theory and dispersion relations,''
  Eur.\ Phys.\ J.\ C {\bf 19}, 517 (2001).
%  doi:10.1007/s100520100629
%  [hep-ph/0012023].
  %%CITATION = doi:10.1007/s100520100629;%%
  %45 citations counted in INSPIRE as of 25 Mar 2016
	




%\cite{GarciaMartin:2011cn}
\bibitem{GarciaMartin:2011cn} 
  R.~Garcia-Martin, R.~Kaminski, J.~R.~Pelaez, J.~Ruiz de Elvira and F.~J.~Yndurain,
  %``The Pion-pion scattering amplitude. IV: Improved analysis with once subtracted Roy-like equations up to 1100 MeV,''
  Phys.\ Rev.\ D {\bf 83}, 074004 (2011).
%  [arXiv:1102.2183 [hep-ph]].
  %%CITATION = ARXIV:1102.2183;%%
  %132 citations counted in INSPIRE as of 23 sept. 2015
  R.~Kaminski, J.~R.~Pelaez and F.~J.~Yndurain,
  %``The pion-pion scattering amplitude. II. Improved analysis above bar K anti-K threshold,''
  Phys.\ Rev.\ D {\bf 74} (2006) 014001
   [Phys.\ Rev.\ D {\bf 74} (2006) 079903].
%  doi:10.1103/PhysRevD.74.014001, 10.1103/PhysRevD.74.079903
%  [hep-ph/0603170].
  %%CITATION = doi:10.1103/PhysRevD.74.014001, 10.1103/PhysRevD.74.079903;%%
  %54 citations counted in INSPIRE as of 18 Dec 2015
  J.~R.~Pelaez and F.~J.~Yndurain,
  %``The Pion-pion scattering amplitude,''
  Phys.\ Rev.\ D {\bf 71}, 074016 (2005).
%  [hep-ph/0411334].
  %%CITATION = HEP-PH/0411334;%%
  %95 citations counted in INSPIRE as of 23 sept. 2015


	

\end{thebibliography}
\end{document}